\documentclass[9pt,twocolumn,twoside]{pnas-new}
\templatetype{pnasresearcharticle} % Choose template 
\usepackage{amsmath}
\usepackage{amsmath}
\usepackage{upgreek}
\usepackage{gensymb}
\usepackage{siunitx}
\usepackage[utf8]{inputenc}

\renewcommand{\phi}{ \varphi }

\newcommand{\degC}[1]{{$^{\rm\circ}$}}
\newcommand{\RomanNumeral}[1]{\MakeUppercase{{\fontfamily{ptm}\selectfont
\romannumeral #1}}}

\title{Deformation Dynamics of Nanopores \\upon Water Imbibition}
\author[a]{Juan Sanchez}
\author[a,b,e]{Lars Dammann}
\author[a,c]{Laura Gallardo}
\author[a,b]{Zhuoqing Li}
\author[d]{Michael Fröba}
\author[e,f]{Robert H. Mei{\ss}ner}
\author[g]{Howard A. Stone}
\author[a,b,f]{Patrick Huber}

\affil[a]{Institute for Materials and X-ray Physics, Hamburg University of
Technology, Hamburg 21073, Germany}
\affil[b]{Center for X-ray and Nano Science, Deutsches Elektronen-Synchrotron, Hamburg 22607, Germany;}
\affil[c]{Centre for the Study of Manuscript Cultures, Hamburg University, Hamburg 20354, Germany;}
\affil[d]{Institute of Inorganic and Applied Chemistry, University of Hamburg, Hamburg 20146, Germany;}
\affil[e]{Institute of soft Matter Modeling, Hamburg University of Technology, Hamburg 21073, Germany}
\affil[f]{Institute of Surface Science, Helmholtz-Zentrum Hereon, Geesthacht 21502, Germany}  
\affil[g]{Department of Mechanical and Aerospace Engineering, Princeton University, Princeton, NJ 08540}

\leadauthor{Sanchez, Dammann} 

\significancestatement{Capillarity-driven flows in nanometer-sized pores play a dominant role in many natural and technological processes, ranging from water transport and transpiration in trees, clay swelling and catalysis to transport through microfluidic structures and fabrication of battery materials. Here we show by a combination of experiments and computer simulations of water imbibition in nanopores that the competition between expansive, surface stress release at pore walls and negative, contractile Laplace pressures of nanoscale menisci lead to an unusual macroscopic behavior of the porous medium, which is generic for any liquid/nanoporous solid combination. The results allow one to quantify surface and Laplace stresses and to monitor nanoscale flow and infiltration states by relatively simple length measurements of the porous medium.}

\authorcontributions{
J.S., Z. L. and P.H. conceived the experiments. J.S. and Z.L. performed the gravimetric, optical and dilatometry experiments. J.S. and L.G performed the later experiments including evaporation effects. L.D. performed the Molecular Dynamics simulations. J.S., L.D., M.F., R.M., H.A.S., and P.H. interpreted the experimental and computer simulation data. J.S., L.D., H.A.S., P.H. wrote the manuscript. All authors proofread the manuscript.}
\authordeclaration{The authors declare no competing interests.}
\equalauthors{\textsuperscript{1}J.S.(Author One) and L.D. (Author Two) contributed equally to this work.}
\correspondingauthor{\textsuperscript{2}To whom correspondence should be addressed. E-mail: patrick.huber@tuhh.de}


\begin{abstract}
Capillarity-driven transport in nanoporous solids is widespread in nature and crucial for modern liquid-infused engineering materials. During imbibition, curved menisci driven by high negative Laplace pressures exert an enormous contractile load on the porous matrix. Due to the challenge of simultaneously monitoring imbibition and deformation with high spatial resolution, the resulting coupling of solid elasticity to liquid capillarity has remained largely unexplored. Here, we study water imbibition in mesoporous silica using optical imaging, gravimetry, and high-resolution dilatometry. In contrast to an expected Laplace pressure-induced contraction, we find a square-root-of-time expansion and an additional abrupt length increase when the menisci reach the top surface. The final expansion is absent when we stop the imbibition front inside the porous medium in a dynamic imbibition-evaporation equilibrium, as is typical for transpiration-driven hydraulic transport in plants, especially in trees. These peculiar deformation behaviors are validated by single-nanopore molecular dynamics simulations and described by a continuum model that highlights the importance of expansive surface stresses at the pore walls (Bangham effect) and the buildup or release of contractile Laplace pressures as menisci collectively advance, arrest, or disappear. Our model suggests that these observations apply to any imbibition process in nanopores, regardless of the liquid/solid combination, and that the Laplace contribution upon imbibition is precisely half that of vapor sorption, due to the linear pressure drop associated with viscous flow. Thus, simple deformation measurements can be used to quantify surface stresses and Laplace pressures or transport in a wide variety of natural and artificial porous media.
\end{abstract}

\dates{This manuscript was compiled on \today}
\begin{document}

\maketitle
\thispagestyle{firststyle}
\ifthenelse{\boolean{shortarticle}}{\ifthenelse{\boolean{singlecolumn}}{\abscontentformatted}{\abscontent}}{}

Liquid-infused nanoporous solids play an increasingly important role as functional materials. One can find them as soft-hard hybrids with superior properties in terms of adaptive wettability \cite{Yao2013}, mechanical actuation \cite{Wang2018FerrofluidInfused, Brinker2022}, adjustable photonics \cite{Sentker2019, Cencha2020}, in nanofluidics \cite{Emmerich2022}, and in energy storage and energy conversion \cite{Simon2008} applications. Also, the development of structural materials that are capable of adapting to changing environmental conditions can be achieved by combining soft, dynamic liquid phases with static, solid phases that act as mechanically robust scaffold structures. The resulting hybrid materials have demonstrated unprecedented properties of stability, adaptability, and stimuli-responsiveness, as desired in many applications such as 3D printing, soft robotics, omniphobic surfaces, microfluidics, and multiphase separation \cite{Zhang2021a, Wong2011, Yao2013}.

It has also long been known, starting with the pioneering work of Meehan \cite{Meehan1927expansioncharcoal} and Bangham and Fakhoury \cite{Bangham1928}, that stresses and strains are generated in porous solids during the adsorption and desorption of gases and liquids \cite{Scherer1986DilatationGlass,Prass2009, Gor2010,Schappert2014VycorDeformation, Balzer2014, Grosman2015, Gor2017, Chen2019SorptionDeformation, Yang2020, Harrellson2023, Gor2024}. Two effects play a major role in the resulting sorption-induced deformation \cite{Gor2015,Gor2017}. First, the adsorption of matter onto the solid pore wall surface typically leads to a reduction in surface stress. The reduced stress results in an expansion of the porous solid (Bangham effect) \cite{Gor2016RevisitingStress}. Second, the formation of menisci in the pore space during capillary condensation can create a negative pressure in the liquid, the magnitude of which is inversely proportional to the pore size. The effect of the negative pressure on the solid surface causes the solid to contract (Laplace effect) \cite{Gor2015ElasticPores}. 

Similar to fluid adsorption processes, deformation of porous solids is also expected to occur in capillarity-driven spontaneous imbibition experiments with wetting liquids \cite{duprat_aristoff_stone_2011, Scherer1986DilatationGlass, Weijs2013Elasto-capillaritySolids} as well as in forced intrusion of non-wetting liquids \cite{Tortora2021, Michel2022DeformationHydrophobicMedium}. 
In fact, studies highlight the importance of Laplace pressure-induced pore space deformations and their impact on imbibition dynamics for soft (macro)porous media such as cellulose sponges \cite{Hoberg2014, Ha2018Poro-elasto-capillarySponges, Siddique2009CapRiseDefPorousMedium}. However, the influence of surface stress on the deformation behavior, which is particularly important for nanoporous media with their large internal surface-to-volume ratio and which is predicted to lead to pore space expansion, has been little studied. Thus, in contrast to gas adsorption-induced deformations, a comprehensive understanding of imbibition-induced deformations as a result of the interplay of Bangham and Laplace effects is still lacking for the important class of liquid-infused nanoporous materials.

Here, we aim to achieve such a mechanistic understanding of imbibition-induced deformation in nanoporous solids by performing combined experiments on water imbibition and the corresponding deformation dynamics of a model nanoporous solid (monolithic Vycor glass). Furthermore, by comparison with atomistic simulations of water imbibition on a single silica nanopore, we trace the distinct deformation regimes observed at the macroscopic scale to the single-pore behavior.

\section*{Results and Discussion}

\section*{Results and Discussion}

\begin{figure*}[!ht]
	\includegraphics[width=1\textwidth]{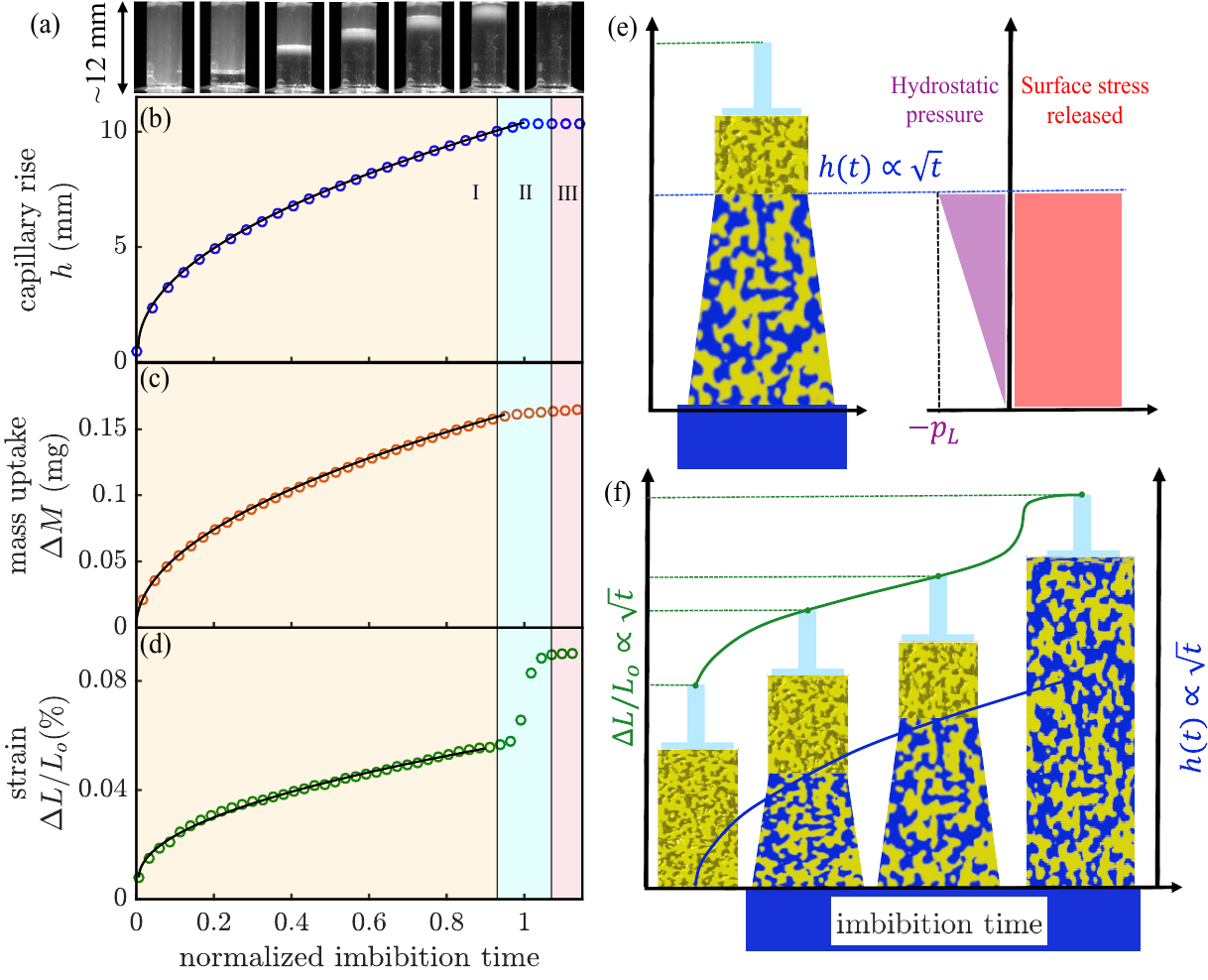}		
	\caption{\textbf{Elastocapillarity dynamics upon water imbibition in nanoporous Vycor glass on the porous-medium scale.} (a) Optical snapshots of water imbibition in a rod-like Vycor monolith. The time between snapshots is 15\,minutes. (b) Mean front position, (c) mass of water uptake, and (d) sample strain measured as a function of time by optical imaging, gravimetry, and dilatometry, respectively. (e) Schematic illustration of the macroscopic deformation of the glass monolith as a function of sample height (left), caused by the interplay of sample contraction, negative hydrostatic pressure (purple) and sample expansion due to surface stress release at the water-silica wall interface per unit sample length (red) as a function of sample height for one selected time $t$ during imbibition. The strain sensor or pushrod is represented as a cylindrical glass piece in contact with the top face of the sample, while the bottom face is in contact with a liquid reservoir. (f) Schematic deformation states for selected times during imbibition (regime \RomanNumeral{1}) and for complete filling (regime \RomanNumeral{3}); also indicated are the resulting changes in the sample length (green line) as well as the imbibition front position (blue line).}
	\label{fig:VycorElastoCapillarityExperiment}
\end{figure*}

\subsection*{Imbibition Experiments with Water: The Interplay of Front Movement, Front Broadening, and Deformation Dynamics}

We study the deformation of nanoporous Vycor glass upon infiltration of liquid water. Vycor is an almost pure fused silica glass permeated by a 3D network of interconnected pores \cite{Levitz1991, Huber1999}. It is formed by a leaching process after spinodal decomposition of a borosilicate glass. Therefore, its geometric structure can be well represented by clipped Gaussian random fields  \cite{Levitz1998, Gommes2018StochasticModelsPorousMedia}. 

To monitor the capillarity-driven spontaneous water imbibition, we perform optical imaging experiments. They allow us to determine the mean front position and its broadening, respectively, due to a transparency change induced by the capillary filling of the pore space and light scattering typical of the inhomogeneously filled pores near the imbibition front, see movie S1\cite{Gruener2009CapillaryNanopores,Gruener2012, Gruener2015, Page1993}. A set of selected snapshots is shown in Fig.~\ref{fig:VycorElastoCapillarityExperiment}a aligned with the optically determined mean imbibition front position (Fig.~\ref{fig:VycorElastoCapillarityExperiment}b), mass uptake of the porous monolith (Fig.~\ref{fig:VycorElastoCapillarityExperiment}c), and the measured sample strain as a function of time (Fig.~\ref{fig:VycorElastoCapillarityExperiment}d).

Since the relative humidity ($RH$) and thus potential adsorption of water via the vapor phase before and during the infiltration experiments is expected to affect both the deformation and imbibition dynamics \cite{Amberg1952b, Gruener2009CapillaryNanopores}, we perform the experiments in a closed cell with controlled humidity.
In the following we first discuss the experiment with a relative humidity ($RH$) of 50\%. According to a water vapor sorption isotherm this corresponds to approximately two pre-adsorbed monolayers of water molecules at the silica pore walls \cite{Gruener2009CapillaryNanopores}. 

Evaporation from the porous block was prevented by shielding the side surface of the Vycor monolith, leaving the top and bottom surfaces unshielded to allow unaltered contact with the dilatometer pushrod and liquid water inlet, respectively. Notice that this configuration allows for air escape through the top facet. Therefore we do not expect any air-entrapment-induced impact on the imbibition and deformation dynamics as has been observed for dead-end single nanopores \cite{Cencha2019}. After initiating capillary rise in the porous glass by raising the level of the bulk liquid reservoir until it reaches the bottom of the glass cylinder, we observe deformation kinetics that include three distinct regimes. Regime \RomanNumeral{1} lasts from the moment of contact with the bulk liquid reservoir until the imbibition front reaches the top of the matrix, see Fig.~\ref{fig:VycorElastoCapillarityExperiment}d. Here the sample expands in accordance with the mean front motion according to a square-root-of-time dynamics. This regime is terminated by a relatively abrupt, additional expansion of the matrix that occurs when the advancing part of the white imbibition front (see Fig.~\ref{fig:VycorElastoCapillarityExperiment}a) reaches the top of the matrix. This step-like deformation continues until the receding part of the white front reaches the top surface, and thus until complete filling of the pore space is achieved. This can be seen in both the optical imaging (see Fig.~\ref{fig:VycorElastoCapillarityExperiment}b) and mass uptake (see Fig.~\ref{fig:VycorElastoCapillarityExperiment}c) experiments. In agreement with a constant final mass uptake and the complete disappearance of the white front, the subsequent regime \RomanNumeral{3} after this ``strain jump'' indicates the complete filling of the pore space and is characterized by a deformation plateau. The total strain measured after complete filling of the matrix is \SI{0.09}{\percent}, which corresponds to a length increase of \SI{10.79}{\micro\metre} of our sample (original length \SI{11.99}{\milli\metre}).

Next, we propose a model for the time-dependent strain evolution. The model requires a macroscopic description of the imbibition dynamics, as well as a characterization of the strain-inducing elastocapillary phenomena occurring at the single-nanopore scale and how they translate to the macroscopic porous-medium scale. \\

\subsection*{Elastocapillarity Dynamics: From the Single-Nanopore to the Porous-Medium Scale}
he imbibition dynamics of the Lucas-Washburn front motion can be derived from a Darcy description of the spontaneous imbibition process \cite{Gruener2009CapillaryNanopores, Gruener2011}. Since at these scales inertial and gravitational effects are negligible compared to capillarity, the imbibition kinetics are governed by a dynamic equilibrium between the approximately constant Laplace pressure at the highly curved menisci of the imbibition front, which drives the flow, and the increasing viscous drag in the liquid-saturated part of the sample, which opposes the flow. As outlined in Ref.~\cite{Gruener2009CapillaryNanopores}, for pore-space geometries such as those present in Vycor glass, the following equation is obtained for the imbibition height $h(t)$ as a function of time $t$:
 \begin{equation}
    \label{eqn:MLW}
   h(t)=\sqrt{\frac{\gamma}{2 \eta \tau  \xi}\frac{(r_p-d)^3}{r_p^2}}\sqrt{t},
\end{equation}
where $r_p$ is the pore radius, $\gamma$ is the water-vapor surface tension, $\eta$ is the viscosity of water, and $\tau$ is the hydraulic tortuosity. The pre-adsorbed water film on the pore walls has a thickness $d$. This pore wall coverage results in a reduced, initial porosity $\Phi_{\rm i}$ accessible to the imbibing water, compared to the original host material porosity $\Phi_0$, measured in vacuum. Therefore, a scaling factor $\xi=\frac{\Phi_{\rm i}}{\Phi_{\rm 0}}$ must be included in Eq.~\ref{eqn:MLW}. We will refer to this equation as the modified Lucas Washburn equation (MLW). 

Eq.~\ref{eqn:MLW} can be extended to describe the mass uptake of water by multiplying the host material area $A$, the reduced porosity $\Phi_{\rm i}$, and the water density $\rho$, giving $\Delta m(t) = \rho \Phi_{\rm i} A h(t)$. Assuming two immobile water layers ($d=$ \SI{0.5}{\nano\metre}), one obtains an excellent fit to the observed mean imbibition position and mass uptake, see Fig.~\ref{fig:VycorElastoCapillarityExperiment}b,c. This reduction in the hydraulic radius can be attributed to the strongly bound, highly viscous interfacial water layers at the silica surfaces, in agreement with previous experiments and computer simulations \cite{Schlaich2017HydrationFriction, Gruener2009CapillaryNanopores}.
Note, that the pre-adsorbed water layers lead to an initial Bangham effect, i.e., adsorption-induced expansion of the Vycor glass relative to the completely dry sample, even before the imbibition experiment is started.

Upon contact with the bulk reservoir, two processes occur in parallel. In the part of the pore space filled with liquid, i.e., the part of the sample up to the capillary rise height $h(t)$, the Bangham expansion takes place due to the corresponding release of surface stress. In addition, the pressure in the liquid increases linearly from the tensile Laplace pressure of the menisci in the advancing imbibition front to the atmospheric pressure at the bulk water reservoir. This results in a contractile stress contribution that decreases linearly from the imbibition front to the bulk reservoir. Thus, the total deformation of the matrix must be considered as a superposition of two opposing deformation effects, the expansive surface stress release (Bangham effect) and the contractile Laplace pressure effect, see Fig.~\ref{fig:VycorElastoCapillarityExperiment}e.

Since the induced strain and thus pore size changes are small, on the order of fractions of a percent, the resulting time-dependent changes in the hydraulic permeabilities and thus capillarity-driven flow dynamics due to the changes in the pore-sizes are negligible. Therefore, both deformation effects scale, to good approximation, linearly with the capillary-filling dynamics. More precisely, the strain dynamics follow the classical square-root-of-time LW dynamics, as proven by our experimental strain observation in regime \RomanNumeral{1}, see Fig.~\ref{fig:VycorElastoCapillarityExperiment}d. 

Since we observe the matrix to be expanding, the expanding Bangham contribution obviously dominates the Laplace pressure effect. The abrupt expansion observed in regime \RomanNumeral{2} is explained by the sudden vanishing of the contractile Laplace pressure contribution. Upon reaching the top surface of the matrix the highly curved menisci and the corresponding Laplace pressure contribution vanish. The disappearance of the contractile contribution leads to an additional sample expansion, which we refer to as the ``Laplace expansion jump.'' The duration of the observed expansion is dictated by the broadening of the imbibition front upon complete filling of the sample and thus by the time characteristic of its vanishing, when the front reaches the top of the sample. This can be seen by a comparison of the vanishing of the white scattering front in the optical images with the times characteristic of the beginning and end of deformation regime \RomanNumeral{2}, respectively (Fig.~\ref{fig:VycorElastoCapillarityExperiment}).

\begin{figure*}[!ht]
	\includegraphics[width=1\textwidth]{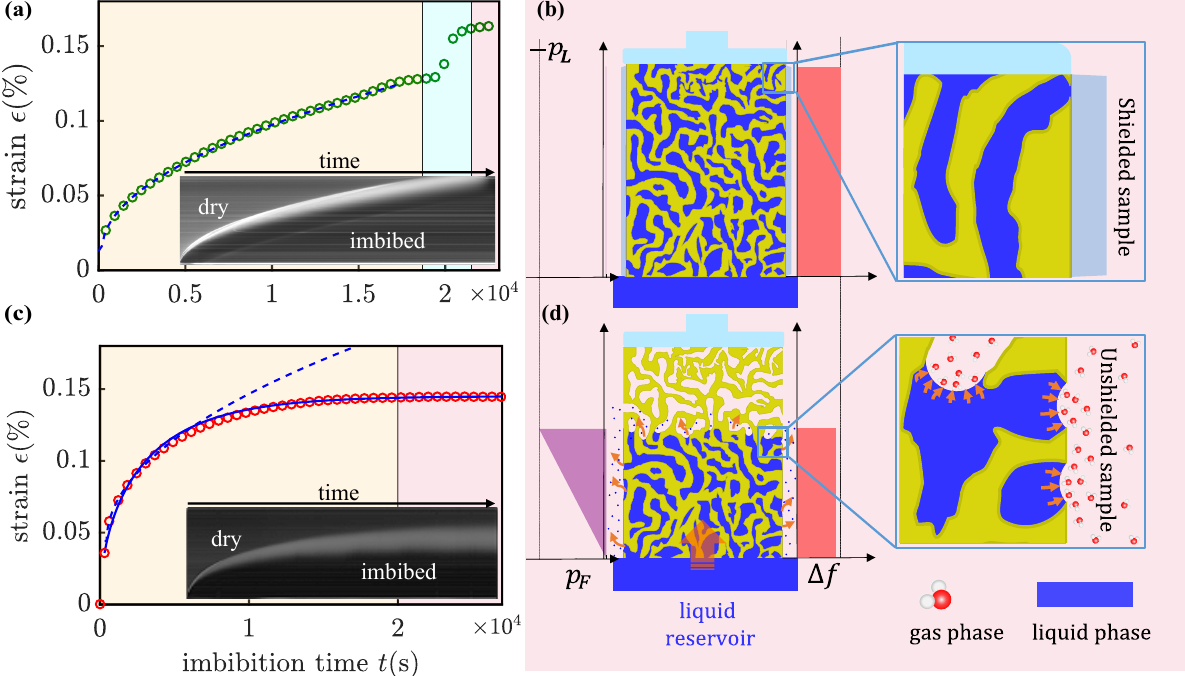}		
	\caption{\textbf{Impact of evaporation on imbibition-induced deformation.} {Imbibition-induced deformation of a nanoporous Vycor glass monolith ($r_p=$ \SI{3.4}{\nano\metre} and $RH\approx$ \SI{5}{\percent}) measured by \textit{in situ} dilatometry in two different scenarios: the first without evaporation (NEv) in (a-b) and the second with evaporation (Ev) in (c-d). In both cases we present our dilatometry data (NEv: green circles; EV: red circles), combined with optical imaging. We select a vertical pixel line centered in the sample and integrate over time. The visible white region reveals the broadening of the imbibition front separating the imbibed and dry portions of the porous host. The dashed lines in (a) and (c) correspond to fitting curves based on a $\sqrt{t}$ scaling. The fitting curves in (c) do not start at $t=0$ because they do not account for the early times when unavoidable lateral imbibition resulted from the insertion of the bottom of the unsealed into the liquid reservoir. The deformation dynamics are compatible with the expected exponential relaxation of the imbibition process, supported by the blue solid fitting line (see Supporting Information for derivation). In the non-evaporation scenario, such a condition is achieved by sealing the monolith lateral surface as shown in the panel (b). Here we show a representative 2-D slice of the completely filled porous matrix in contact with a liquid reservoir. The top of the sample is effectively sealed by the dilatometer pushrod. The orange arrows represent the water flow from the reservoir to the pore space (imbibition). Zooming into the upper region, we see the absence of curved menisci since there is no evaporating outflow to counterbalance the capillarity-driven imbibition. This contrasts with the presence of menisci at the saturated-unsaturated interface, as shown in panel (d). Here we illustrate the dynamic equilibrium reached for an unsealed sample when both the evaporation and imbibition flow rates are equal. The orange arrows at the curved water-gas interface symbolize the evaporation of water molecules.}}
	\label{fig:ImbEvapEquilibrium}   
\end{figure*}

Water vapor transport through the pore network is very slow compared to capillarity-driven liquid flow, so the effects of evaporation through the top surface during imbibition are negligible. Additionally, the upper dilatometer glass pushrod effectively sealed the upper sample surface. In the absence of a mechanism to counteract water uptake, the liquid will continue to rise according to MLW dynamics until the pore space is completely filled \cite{Gruener2009CapillaryNanopores, Maillet2022}. In Fig. \ref{fig:ImbEvapEquilibrium} (top) we show such dynamics and qualitatively illustrate the flattening of the menisci once the liquid reaches the upper porous block surface and thus the upper monolithic glass pushrod.

\subsection*{Impact of Evaporation on Elastocapillarity Dynamics: The Extreme Case of Reaching an Evaporation-Imbibition Equilibrium}
We performed a second imbibition experiment in which, unlike the first, the side facets are not sealed against evaporation. Then, the evaporation fluxes increase in proportion to the wetted, vapor-phase exposed part of the side facets and thus in proportion to the capillary rise height. The imbibition front then stops at a height well below the upper monolith surface as the capillary rise reaches dynamic equilibrium, see movie S2 in the supporting material. The inflow from the bulk liquid reservoir equals the evaporative outflow to the bulk vapor phase, so that an evaporation-driven water transport reminiscent of an (artificial) tree is achieved \cite{Seker2008WettingEvaporation, Wheeler2008, Camplisson2015}, see Fig.~\ref{fig:ImbEvapEquilibrium}a.  

In this case, the dynamics of capillary rise deviates from the classical $\sqrt{t}$ scaling. As shown in Fig. \ref{fig:ImbEvapEquilibrium} (bottom, left), the water reaches a final capillary rise height, where the arrested imbibition front consists of highly curved menisci. From the imbibition front to the water reservoir there exists a linear pressure gradient that continues to exert a tensile contraction on the wetted porous material. Therefore, no ``Laplace expansion jump'' can be observed in the sample even though the imbibition front stopped rising.

As outlined in the Supplementary Information the capillary-rise dynamics for this scenario can be described by 
\begin{equation}
\label{eqn:exp_relax}
    h(t)=h_{\rm eq}\cdot \sqrt{1-e^{-\frac{4q}{R \Phi_{\rm i}}t}},
\end{equation}where $q$ is the evaporation rate of water per surface unit and $h_{\rm eq}=\sqrt{\frac{K \Delta p R}{2q \mu }}$ represents the equilibrium capillary rise or saturation height, reached at sufficiently long times. The hydraulic permeability $K$ and every variable involved in $h_{\rm eq}$ except for $q$ can be obtained for the case of water imbibition in Vycor glass from Refs.~\cite{Gruener2009CapillaryNanopores, Maillet2022}.    

Since the deformation dynamics evolve proportionally to the capillary rise, the inclusion of a proportionality factor $C$ allows
the use of Eq.~\ref{eqn:exp_relax} as a fitting function, $\epsilon(t)=C \cdot \sqrt{1-e^{-\beta t}}$, for the dilatometry data. Our experiment leads to $\beta=$ \SI{2.1e-4}{\per\second}, which translates to an evaporation rate $q=$ \SI{0.05}{\micro\metre\per\second}. This value is in reasonable agreement with values reported in literature $q\approx$ \SI{0.045}{\micro\metre\per\second} for bulk deionized water at similar conditions of humidity and temperature \cite{Huang2021}. The determined evaporation rate $q$ leads to an equilibrium capillary rise of $h_{\rm eq}=$ \SI{14}{\milli\metre} or a filling height of \SI{74}{\percent} of the total length (\SI{19}{\milli\metre}), which is in reasonable agreement with the observed averaged position of the arrested front visible in the snapshots in Fig.~\ref{fig:ImbEvapEquilibrium}c.

Note that a first-order Taylor series expansion of Eq.~ \ref{eqn:exp_relax} around $t=0$ leads back to our MLW law or Eq.~\ref{eqn:MLW}. This means that for early times ($t\ll t_{\rm E}=\frac{R\phi_{\rm i}}{4q}$), where $t_{\rm E}= \SI{4700}{\second}\pm{\SI{600}{\second}}$ corresponds to the time when the evaporative flow rate has reached 1-1/e=63\% of the imbibition flow rate, imbibition dominates over evaporation and the rising dynamics can be described exclusively in terms of the classical LW square-root-of-time dependent law (see Supplementary Information). This is evidenced by the overlapping of the fitting curves in Fig.~ \ref{fig:ImbEvapEquilibrium}c, for $t<t_{\rm E}$, where the continuous line indicates the fitting model based on Eq.~\ref{eqn:exp_relax} and the dashed line denotes the $\sqrt t$-fitting based on Eq.~ \ref{eqn:MLW}. The latter is in reasonable agreement with the deformation dynamics shown in Fig. \ref{fig:ImbEvapEquilibrium} a, corresponding to a non-evaporation scenario (sealed sample), and is very well described by the MLW model. The fitting coefficients are $C_{\rm unsealed}=$\SI[separate-uncertainty = true]{8.53(50)e-4}{\per\second\tothe{1/2}} and $C_{\rm sealed}=$ \SI[separate-uncertainty = true]{10.4(2)e-4}{\per\second\tothe{1/2}}.

In the derivation of Eq.~\ref{eqn:exp_relax} we consider only the liquid flow in the vertical imbibition direction, so that a one-dimensional (1-D) problem results for $h=h(t)$ and no transversal imbibition front height variation is considered. However, a closer look at the shape of the imbibition front reveals that it becomes increasingly curved with increasing imbibition time, with $h$ being maximal in the center of the sample and decreasing symmetrically in the radial direction towards the sample sides, as evident in the snapshot sequence in Fig.~S1 of the Supplementary. This suggests an increasing importance of the transversal flow components in the imbibition phenomenology with increasing imbibition time. In fact, the slight but systematic overestimation of the strain by our simple 1-D model compared to the measured strain in the intermediate time range from about \SI{5000}{\second} to the onset of saturation at about \SI{15000}{\second} remains unexplained. Perhaps it results from neglecting the curvature of the imbibition front and thus the lateral flow components, since the average height of the curved front and thus the averaged Bangham effect is smaller than expected in the simplified model with a flat imbibition front. Furthermore, we neglect the Kelvin effect, i.e. a reduction of the vapor pressure at the highly curved menisci, which could also lead to an underestimation of the equilibrium height $h_{\rm eq}$, since it results in reduced evaporative fluxes compared to the planar bulk interface. 

\subsection*{Quantitative Modeling of the Deformation Dynamics on the Porous-Medium Scale}
To achieve a quantitative description of the deformation kinetics we consider the monolithic porous glass as a 3-D network of interconnected cylindrical channels with an isotropic orientation distribution. We neglect the broadening of the imbibition front and therefore assume that the imbibed part of the network is completely filled with liquid up to the height $h(t)$. 

As outlined in detail in Ref.~\cite{Gor2016RevisitingStress}, given the narrow pore-size distribution of Vycor glass and the disordered channel-like structure of the pores, its deformation can be considered uniform, isotropic, and a single pore-size approximation can be used. The deformations are determined by a surface stress contribution and by the hydrostatic fluid pressure in the pore $p_{\rm \rm F}$, both acting as an effective loading pressure on the solid and normal to the pore walls. These contributions result in the equations:
 
\begin{equation}
\label{eqn:modelP0}
    \begin{split}
        \epsilon_\mathrm{Bangham} = -\frac{1}{M_\mathrm{PL}} \frac{\Delta f}{r} \quad\hbox{and}\quad
        \epsilon_\mathrm{Laplace} = \frac{1}{M_\mathrm{PL}} p_{\rm F},
    \end{split}
\end{equation}
where $M_\mathrm{PL}$ is the hydrostatic pressure-associated pore-load modulus \cite{Prass2009}, $\Delta f$ is the released surface stress, i.e., the difference in surface stress before and after imbibition, and $r=r_p-d$ is the corrected pore radius, which appears in Eq. \ref{eqn:MLW}.

For isotropically distributed cylindrical pores, the total strain of a liquid-filled representative volume $\epsilon^*$ is calculated then as a combination of the two strains:
\begin{equation}
\label{eqn:modelP}
    \begin{split}
        \epsilon^* &= \frac{1}{M_\mathrm{PL}} \left(-\frac{\Delta f}{r}+p_{\rm F} \right).
    \end{split}
\end{equation}

On the porous-medium scale, the fluid pressure increases linearly from the negative, contractile Laplace pressure $p_{\rm L}$ at the position of the advancing imbibition front $h(t)$ up to the bulk reservoir pressure $p_{\rm 0}$, while the change in surface stress per unit length for the imbibed part of the pore network is constant along the sample height at any imbibition time $t$. Thus the Laplace pressure-induced deformation is maximal at the imbibition front and vanishes at the bulk reservoir. By contrast, the Bangham effect is constant along the imbibed part of the matrix, so that the water-filled part of the originally dry, cylindrical monolith deforms to a frustum (see Fig. \ref{fig:VycorElastoCapillarityExperiment}e for the corresponding trapezoidal deformation in side view). The resulting overall deformation and thus relative sample length change $\Delta L/L_{\rm 0}$ in the imbibition direction is thus proportional to the fraction of imbibed sample ($h(t)/h_{\rm final}$). And because of the linearly increasing magnitude of the fluid pressure from the bulk reservoir to the imbibition front, it is only half as large as it would be for a porous solid homogeneously experiencing the tensile pressure $p_{\rm L}$. Additionally considering that for nanoporous materials the atmospheric pressure is much smaller than the Laplace pressure, $p_{\rm 0} \ll p_{\rm L}$, we arrive at:
\begin{equation}
\label{eqn:model2}
    \begin{split}
        \epsilon(t) &= -\frac{1}{M_\mathrm{PL}} \left(\frac{\gamma +\Delta f}{r}\right)\frac{h(t)}{h_{\rm final}}.\\
    \end{split}
\end{equation}

At imbibition time $t_c$ the average position of the imbibition front is equal to the height of the sample $L$ or, in other words, using Eq.~\ref{eqn:MLW}, $h(t_c)= L$. For a single nanocapillary, the tensile Laplace pressure vanishes suddenly at $t_c$, but for our porous matrix such a vanishing is not instantaneous but gradual, since the vanishing dynamics is governed by the broadened imbibition front. The Laplace pressure contribution starts to decrease when the upper limit of the broadening front reaches
 $h(t_1)=L$ and continues to do so until its disappearance when the whole matrix is filled at $t_2$. This time difference $t_2-t_1$ over which the Laplace pressure vanishes is indicated in all of our figures as regime II (light blue background). The extent of the Laplace jump is:
\begin{equation}
\label{eqn:jump}
    \begin{split}
        \Delta\epsilon = \epsilon(t_2)-\epsilon(t_1) =\frac{p_F}{M_\mathrm{P_L}}=\\
        =-\frac{p_L}{2 M_\mathrm{PL}}=-\frac{\gamma}{(r_p-d) M_\mathrm{PL}}
    \end{split}
\end{equation}

\begin{figure*}[!ht]
	\includegraphics[width=1\textwidth]{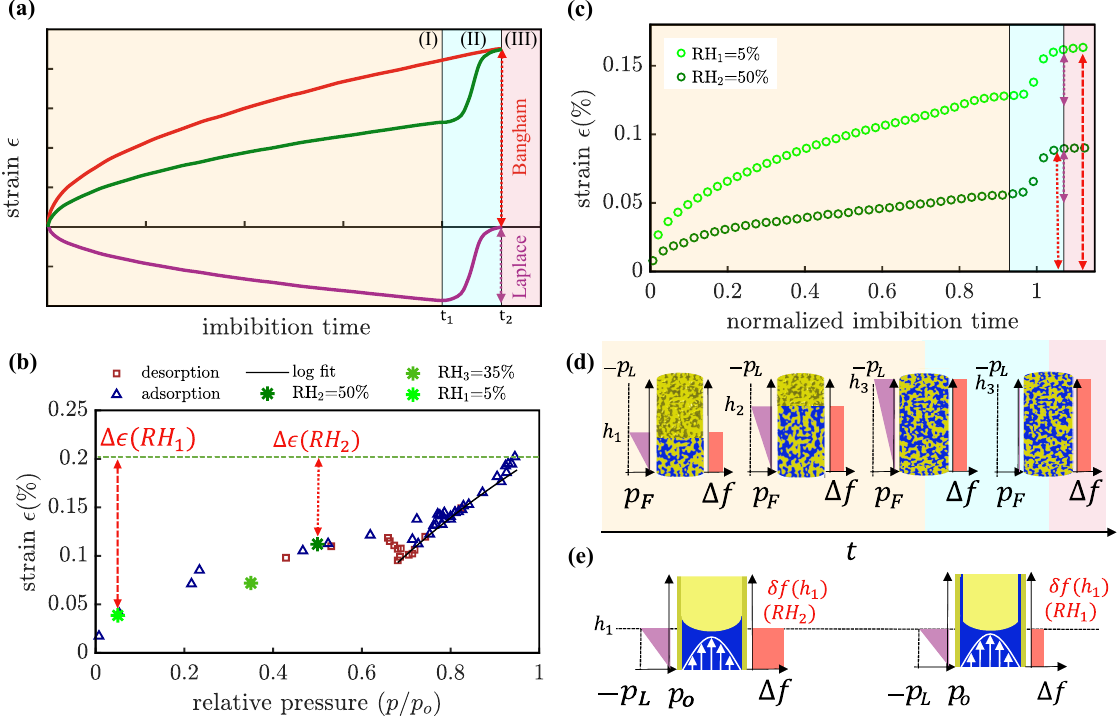}		
	\caption{\textbf{Contributions to imbibition-induced deformation and influence of pre-imbibition humidity.}  (a) Contributions to deformation of the porous-medium  along the imbibition direction that originate from contractile Laplace pressure (purple) and expansive surface stress contributions (red) along with the resulting overall deformation (green line), as calculated by a linear superposition according to our elastocapillary model. The maximum values of both contributions are represented by the dotted lines. (b) Strain isotherm measurement for a Vycor glass rod $r_p=$ \SI{3.4}{\nano\metre}, $T=$ \SI{25.8}{\celsius} and $p_0=$ \SI{3.3}{\kilo\pascal}, with the logarithmic vanishing of the Laplace pressure highlighted with a black line (adapted from Ref. ~\cite{Amberg1952}). Overlapping with such data set we include our own experimental data (green points) obtained by subtracting the maximum strain (Bangham contribution) in our imbibition experiments to the maximum strain value in the isotherm (\SI{0.2}{\percent}) represented by the green dashed line. The red dashed lines represent the measured strain in our dilatometry experiment and the position along the Y-axis of the influence of the adsorption-induced strain due to the humidity. (c) Water imbibition-induced deformation in a nanoporous Vycor sample $r_p=$ \SI{3.4}{\nano\metre} with pre-imbibition humidity  $RH_2\approx$ \SI{50}{\percent}) (dark green points) compared to one (light green dots) performed with lower pre-imbibition humidity ($RH_1\approx$ \SI{5}{\percent}). The strain jump and maximum strain are given by the red dashed and dotted lines for $RH_1$ and $RH_2$, respectively. %{\color{blue} can $RH_1$ and $RH_2$ be indicated on the figure?} 
 (d) Schematic illustrations of the interplay of negative hydrostatic pressure (purple) and a sample-expanding surface stress release at the water-silica wall interface per unit of sample length (red) for specific rise heights ($h$): (h$_1$), (h$_2$), (h$_3$) and (h$_4$). Analogously, for a selected time $t_s$, in (e) this the interplay is shown at the single-pore scale for two distinct pre-imbibition humidity values. The white arrows represent the parabolic velocity profile for pressure-driven flow in a cylindrical channel.}
	\label{fig:VycorElastoCapillarityModelling}   
\end{figure*}

The individual contributions to the strain and consequently the total strain are time dependent and evolve proportional to the MLW dynamics of Eq.~\ref{eqn:MLW}.
Fig.~\ref{fig:VycorElastoCapillarityModelling}a shows a model of both strain contributions and, how the combination of both allows reconstructing the observed deformation dynamics represented by the green curve. This figure panel illustrates the competitive deformation process and does not represent an actual experiment. In this way we  understand the imbibition-induced strain curve of water in nanoporous Vycor glass and why, upon filling, we observe a dynamic expansion proportional to $\sqrt {t}$; the green curve acceptably describes the first stage of the process, prior to a strain jump that lasts until time $t_{\rm c}$ of complete filling.

The strain jump in our dilatometry measurement can be quantified according to Eq.~\ref{eqn:jump} provided the pore-load modulus $M_\mathrm{PL}$ of the material is known. $M_\mathrm{PL}$ is extractable through a sorption strain isotherm measurement \cite{Gor2016RevisitingStress}. The water adsorption-desorption-induced strain on nanoporous Vycor rods was experimentally studied in the 1950s by McIntosh and Amberg \cite{Amberg1952}. We work with nanoporous Vycor hosts with comparable pore size and porosity, and at identical temperature. The pore-load modulus of Vycor is accessible via the Kelvin-Laplace equation upon reaching bulk liquid-vapor coexistence. The McIntosh and Amberg data shows the logarithmic vanishing of the Laplace pressure upon approaching bulk liquid coexistence and thus vanishing menisci curvature (see Fig.~\ref{fig:VycorElastoCapillarityModelling}b). It results in $M_\mathrm{PL}=$ \SI{39.45}{\giga\pascal}. Introducing this value to Eq.~\ref{eqn:jump} while assuming the bulk surface tension of water (\SI{72} {\milli\newton\per\metre}) we get a value for the maximum Laplace pressure-induced strain of \SI{0.036}{\percent}, which is in reasonable agreement with the experimental value of \SI{0.032}{\percent} (see Fig.~\ref{fig:VycorElastoCapillarityModelling}c).  Here the final strain value after the jump (dotted red line) corresponds to the Bangham effect contribution to the strain.

An identical experiment is performed for a relative humidity of $\sim5$\% and also shown in Fig. \ref{fig:VycorElastoCapillarityModelling}c. Again, the three different deformation regimes are observed. Note, however, that the total relative expansion after complete filling is \SI{0.16}{\percent}, almost twice as large as in the experiment at $RH=$ \SI{50}{\percent}, while the strain jump at complete filling is the same in both experiments.

Whereas the effect of the Laplace pressure on the strain upon imbibition is predictable, the surface stress contribution is highly dependent on the pre-imbibition atmosphere. Higher humidity values in our experimental environment, prior to the start of the imbibition experiment, will induce larger pre-imbibition adsorption of water molecules on the pore walls, leading to expansion of the porous host (as illustrated in Fig.~\ref{fig:VycorElastoCapillarityModelling}d). Such dependency is evidenced by comparing the  dilatometry experiments at different pre-imbibition humidities. Again in Fig.~\ref{fig:VycorElastoCapillarityModelling}b we show how, for a certain relative humidity pre-condition $RH_1$, the measurable imbibition-induced deformation (c) is comparable to the strain difference between equilibrium states at $RH_1$ and $RH_f=p/p_0=1$ (fully filled sample).

From the final strain values it is possible to calculate $\Delta f$ using Eq.~\ref{eqn:modelP}. We find $\Delta f(RH=50\%)=$ \SI{-116} {\milli\joule\per\metre\squared} and $\Delta f(RH=5\%)=$ \SI{-206} {\milli\joule\per\metre\squared}. Our simple semi-empirical approach to calculate $\Delta f$ from $\epsilon$ is in reasonable agreement with an effective medium-based description of surface stress-induced mechanical deformation as proposed by Weissmüller \textit{et al}. \cite{Weissmuller2010DeformationForces}. For a network of interconnected elongated fibers with circular cross-section they propose
\begin{equation}
   \label{eqn:weissmueller} 
   \epsilon=-\frac{\alpha \Delta f}{3K}\frac{1-\nu}{1-2\nu},
\end{equation}
 with $\alpha$ being the volume-specific surface area and $K$ being the bulk modulus. The $\Delta f$ values obtained in this way are \SI{-116} {\milli\joule\per\metre\squared} and \SI{-190} {\milli\joule\per\metre\squared} for $RH=50\%$ and $RH=5\%$, respectively.

Considering the common case where the surface stress release equals the difference between the surface energy of the dry solid and the wetted solid, $\Delta f = \gamma_{\rm sl}-\gamma_{\rm sv}$ (sl: solid-liquid, sv: solid-vapor)~\cite{Gor2010a}, Eq.~\ref{eqn:modelP} can be written as
\begin{equation}
\label{eqn:epsilon_spreading}
\epsilon^*=-\frac{1}{M_{\rm PL}}\frac{\gamma+\gamma_{\rm sl}-\gamma_{\rm sv}}{r}= \frac{1}{M_{\rm PL}} \frac{S}{r},
\end{equation}
where $S=\gamma_{sv}-(\gamma_{sl}+\gamma)$, is the spreading coefficient at the liquid meniscus contact line of the wetting theory \cite{Gennes2004}. Hence, for the complete wetting case, Eq.~\ref{eqn:epsilon_spreading} always predicts a positive strain and we arrive at the important conclusion that for any capillarity-driven, spontaneous imbibition process and for any liquid/porous solid combination the Bangham effect prevails over the Laplace effect, i.e., in regime \RomanNumeral{1}, an expansive behavior will always be observed. Note that the complete wetting case assumes a zero liquid/solid contact angle $\theta$. The inclusion of a finite contact angle ($\theta>0$) implies the replacement of $\gamma$ by $\gamma\cos{\theta}$ both for the calculation of the Laplace pressure and the spreading coefficient.

Unfortunately, the surface energy values for dry amorphous silica and hydrous amorphous silica interfaces reported in the literature are quite inconsistent, presumably because of the complex water-silica interaction so that they depend sensitively on any contamination, the thickness of the adsorbed water layers, and the surface chemistry (silanol density and hydroxylation of the surface) ~\cite{Brunauer1956, Diaz2022}. However, for the surface energies ($\gamma$) reported in Ref.~\cite{Brunauer1956} ($\gamma_{\rm sl}=$ \SI{130} {\milli\joule\per\metre\squared} and $\gamma_{\rm sv}=$ \SI{260} {\milli\joule\per\metre\squared}, respectively) we arrive at a predicted surface stress release of $\Delta f=$ \SI{-130} {\milli\joule\per\metre\squared} for our $RH=$ \SI{50}{\percent} measurements, in reasonably good agreement with our finding of \SI{-116} {\milli\joule\per\metre\squared}. This also corroborates the conclusion in Ref.~\cite{Gor2016RevisitingStress} that for water-silica interfaces the surface stress corresponds to the surface energy changes.

\subsection*{Molecular Dynamics Simulations of Water Imbibition-Induced Deformation on the Single-Nanopore Scale}

\begin{figure}
	\centering
	\includegraphics{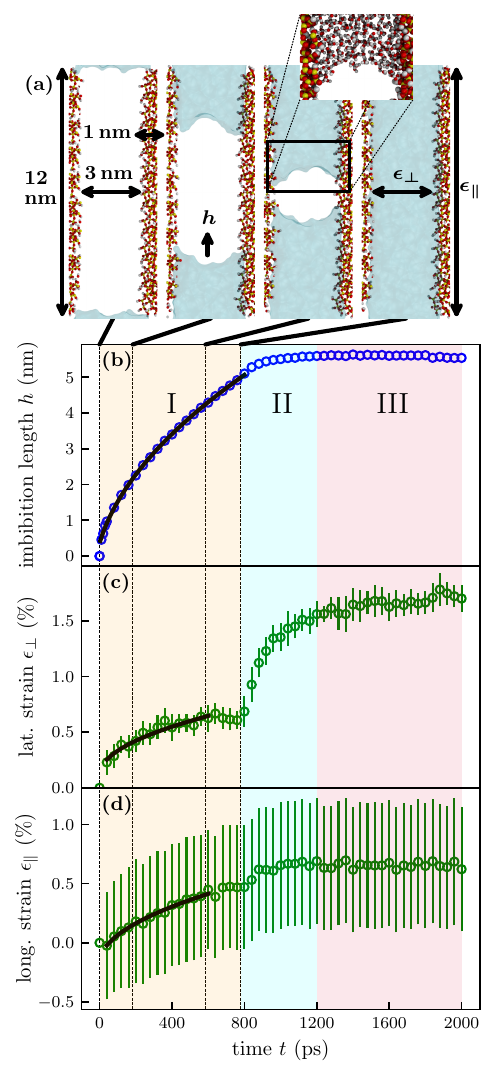}
	\caption{\textbf{Elastocapillary dynamics upon water imbibition on the single-nanopore scale as inferred from molecular dynamics computer simulations.} (a) Snapshots from an exemplary simulation at selected imbibition times as indicated in the figure. Silicon atoms are depicted in yellow, oxygen in red, and hydrogen in white. The imbibing water is represented through the blue colored surface. (b) Imbibition length $h$ of the water column in the silica pores from one pore side in comparison with a $\sqrt{t}$ fit (black). (c) and (d) Observed lateral $\epsilon_{\perp}$ and longitudinal $\epsilon_{\parallel}$ strain in the pore matrix.}
	\label{fig:simulation_results}
\end{figure}
As demonstrated above, dilatometry experiments allow the study of the macroscopic effects of imbibition-induced strain on porous host materials and a quantification of the total strain. However, the individual strain contributions originating from the superposition of the Bangham and Laplace pressure effects on the single-pore scale can not be investigated with this technique.

To better understand how these microscopic strain contributions accumulate to yield the total macroscopically measured deformation and how they are related to individual menisci movements in the pore space, Molecular Dynamics (MD) computer simulations of the imbibition of water into silica nanopores on the single-pore scale are conducted. Since in MD simulations atom-atom interactions for each individual atom in the system are modeled they are well suited to investigate the microscopic imbibition-induced strain contributions on a fundamental level. The  porous silica system used is based on the original work of Refs.~\cite{Ugliengo2008RealisticCalculations, DellePiane2014Large-scaleSystem} and has a diameter of around \SI{3}{\nano\metre} and a wall thickness of around \SI{1}{\nano\metre}. The total length the water can imbibe is about \SI{12}{\nano\metre}. 

Fig.~\ref{fig:simulation_results} shows the results of an ensemble of 20 MD simulations with independent starting conditions; see movie S3 of one of the simulations in Supporting Information. 
Fig.~\ref{fig:simulation_results}a displays snapshots of slices through a nanopore during the imbibition process. The snapshots are arranged from left to right in chronological order at times \SI{0}{\pico\second}, \SI{183}{\pico\second}, \SI{585}{\pico\second} and \SI{780}{\pico\second} as indicated by the dotted vertical lines in figures \ref{fig:simulation_results}b, \ref{fig:simulation_results}c, and \ref{fig:simulation_results}d. In the third snapshot a zoom-in is provided that focuses on the water meniscus formed at the imbibing water column. Note that with a \SI{3}{\nano\metre} pore diameter the silica pore size is only around one order of magnitude larger than the size of a water molecule as visible from the inset.

The imbibition length $h$ of the water column into the silica pore from one pore side is shown as a function of imbibition time $t$ in Fig.~\ref{fig:simulation_results}b. The error bars are omitted since the plot markers are larger than the standard deviation of the simulation ensemble. In the simulations the imbibition length is measured indirectly through the extent of the simulation box in the longitudinal pore direction. The initial imbibition dynamics of around \SI{3} {\pico\second} follows a linear trend, which is consistent with an initial water inertia-governed dynamics during which the influence of the water viscosity can be neglected (see Supporting Information for further information). After the initial \SI{3}{\pico\second} the imbibition dynamics exhibits a $\sqrt{t}$ dependence as shown by the black fit, which confirms the validity of the LW-model even for pore sizes that only fit a few adjacent water molecule layers. These results are in agreement with previous MD simulations of nanopore imbibition for simple Lennard-Jones fluids and polymer melts ~\cite{Dimitrov2007}. The pore is completely filled with water after \SI{780}{\pico\second}. The measurement of the imbibition length through the extent of the simulation box in the longitudinal pore direction assumes a constant pore space volume. For the process of imbibition this is a good approximation since the pore space volume only changes gradually. However, at the end of the imbibition process a relatively strong lateral pore expansion occurs (see Fig.~\ref{fig:simulation_results}c). This lateral expansion leads to a significantly increased pore space volume that triggers an additional water inflow from the water reservoir into the pore space, even though the pore is already entirely filled by the imbibing water column. The inflow of the water in turn reduces the reservoir volume and leads to a shrinkage of the simulation box in the longitudinal pore direction. This effect explains the additional increase of measured imbibition length even after the water columns fill the entire pore space.

The green markers in Fig.~\ref{fig:simulation_results}c and \ref{fig:simulation_results}d display the dynamics of the lateral $\epsilon_{\perp}(t)$ (perpendicular to the long pore axis) and longitudinal $\epsilon_{\parallel}(t)$ (parallel to the long pore axis) strains, respectively. Error bars indicate the standard deviations over the averaged 20 individual simulations. The use of different analysis methods to determine the lateral and the longitudinal strains explains the different magnitudes of standard deviation in the measured strains (see Materials and Methods). The strong changes in strain during the first time steps of the simulation are in response to the initial inertia-dominated imbibition and initial system equilibration dynamics.

The temporal strain evolution, laterally $\epsilon_{\perp}(t)$ and longitudinally $\epsilon_{\parallel}(t)$, can be divided in three regimes. Regime \RomanNumeral{1} from \SI{0}{\pico\second} to about \SI{780}{\pico\second} displays an expansion of the host material with $\sqrt{t}$ dynamics as indicated by the black curve fit. In agreement with our macroscopic experiments on the porous-medium scale, the expansive Bangham effect dominates in comparison to the contractile Laplace pressure effect on the single-pore scale. However, the influence of the Laplace pressure effect is different for the two strain directions. In the lateral direction the negative Laplace pressure leads to a pronounced contractile effect. In contrast, the longitudinal strain is only influenced by the Laplace pressure through the Poisson effect. More precisely, the negative Laplace pressure leads to an expansion in pore wall thickness, which, in turn through the Poisson effect, leads to a contractile effect of small magnitude in the longitudinal pore direction. At the end of regime \RomanNumeral{1} the menisci at the water columns vanished upon complete filling of the pore space, which leads to a disappearance of the contractile Laplace pressure and the resulting strain effects in the lateral and longitudinal directions. Consequently the strain increases during the relaxation of the porous material from \SI{780}{\pico\second} to \SI{1200}{\pico\second} in regime \RomanNumeral{2}. Regime \RomanNumeral{3} occurs after the expansion around \SI{1200}{\pico\second} when the porous material exhibits the final observed strain created by the Bangham effect acting on the entire inner surface of the nanopore.

Thus, overall the strain evolution at the single-pore scale is analogous to that observed at the macroscopic porous-medium scale. In particular, by
 comparison with the single-pore simulation, the Laplace pressure strain jump observed when the imbibition front reaches the top of the Vycor matrix can be attributed to a collective disappearance of menisci in Vycor.

Analogous to the macroscopic case, for the simulated material $M_\mathrm{PL}$ can be calculated from Eq.~\ref{eqn:jump} with the Laplace jump $\Delta\epsilon=$ \SI{1}{\percent}. This results in a pore-load modulus value of $M_\mathrm{PL}=$ \SI{4.8}{\giga\pascal}, an order of magnitude smaller than the \SI{39.45}{\giga\pascal} calculated for Vycor glass. This is due to the significantly smaller silica pore wall thicknesses in the simulation compared to the experiment, and also explains why the magnitude of the imbibition-induced strains is an order of magnitude larger than the experimental one. From the total strain we calculated $\Delta f_{\rm simulation}=\Delta \gamma_{\rm simulation}=$ \SI{-177}{\milli\joule\per\metre\squared} provided that $p_{\mathrm{L}}=(\gamma\cos{\theta})/r_p$ and the contact angle is about $\theta=$ \SI{45}{\degree}, as can be inferred from the meniscus snapshots in Fig~\ref{fig:simulation_results}. We find that this value (starting from vacuum) is the same order of magnitude as the calculated values of $\Delta \gamma$ at low humidity in the macroscopic case (between \SI{-190}{\milli\joule\per\metre\squared} and \SI{-206}{\milli\joule\per\metre\squared}). Thus the single-pore simulations corroborate not only qualitatively, but also (semi-)quantitatively the experimental findings on the porous-medium scale. 

\section*{Summary}
We have studied experimentally and by computer simulation the deformation behavior of a mesoporous silica monolith (Vycor) and a single cylindrical nanopore, respectively, as a function of water imbibition. In contrast to the expected Laplace pressure-induced contraction, we observed two distinct imbibition-induced expansion kinetics, which can be directly attributed to the advance, arrest, and disappearance of the imbibition front or single nanoscale menisci, respectively. In particular, the surface stress release-induced expansion known as the Bangham effect was found to dominate the Laplace pressure-induced contraction in both cases. As we derive from general wetting theory, this is the generic case for any spontaneous imbibition process and any liquid/solid combination, provided that the surface stress release is equal to the surface energy difference between the wetted and dry solid.

Notably, there is also an important difference in the Laplace contribution to the deformation and thus in the total deformation between filling by classical, quasi-static water vapor adsorption and dynamic liquid water imbibition. In the case of gas sorption, the entire filled part is under a homogeneous negative Laplace pressure \cite{Gor2015ElasticPores, Gor2017}. In the case of spontaneous imbibition, the tensile pressure in the porous medium decreases linearly from the advancing menisci to the bulk reservoir due to viscous liquid flow. This results in exactly half the Laplace pressure-induced contraction of the porous medium compared to the situation with gas adsorption (capillary condensation/evaporation). Since the contractile Laplace effect counteracts the dominant expansive Bangham effect, the total deformation is always smaller for gas sorption than for liquid imbibition. Thus, for fragile porous solids, filling by gas sorption should generally be advantageous, since smaller mechanical strains are expected.

The observed competition between surface stress and Laplace pressure should be relevant for any imbibition phenomenon in nanoporous media. For significantly stiff systems, where deformation leads to negligible changes in pore size and thus hydraulic permeability, simple LW deformation kinetics are expected. For soft solids, on the other hand, the deformation could lead to linear hydraulic resistance gradients and thus to deviations of both the liquid advancement kinetics and the deformation dynamics from the classical LW behavior. Also, for extremely narrow, sub-nanometer pores,  repulsive hydration layer forces could have a non-trivial influence on the elastic imbibition response.

Our results show that relative length changes can be used to study the liquid filling kinetics and thus the fabrication of the emerging class of liquid-infused functional materials in detail, but also under in operando conditions a study of the deformation behavior allows for monitoring the appropriate filling state of a liquid-infused materials system. 
 
\showmatmethods{} % Display the Materials and Methods section

\section*{Materials and Methods}
\subsection{Humidity-Dependent Dilatometry Experiments}

As porous host we work with cylindrical monoliths of porous Vycor glass (Corning Glass, 7930). All of the samples are identical, only differing in length, and were previously used by Simon Grüner. An extensive description of such samples can be found in Ref.~\cite{Gruener2009CapillaryNanopores}. Important properties such as the mean pore radius $r_p=$ \SI{4.9}{\nano\metre} and porosity $\Phi_0=0.3$ are obtained from a water sorption isotherm measurement. Ref.~\cite{Gruener2009CapillaryNanopores} assumed that there are two pre-adsorbed monolayers of water molecules of thickness $d=$ \SI{0.5}{\nano\metre} attached to the pore walls prior to the sorption isotherm measurements. Since we measure in comparable atmospheric conditions we will assume the same value for our measurements. This magnitude of water layer thickness is consistent with the reduced porosity $\Phi_0=0.27$ that we estimate by measuring the sample mass difference between the beginning and end of the experiment. The tortuosity value of $\tau=3.6$ is also taken from the above mentioned study. 

The samples are pre-cleaned by immersing them in a mixture of concentrated sulfuric acid (H\textsubscript{2}SO\textsubscript{4}) and a \SI{30}{\percent} hydrogen peroxide solution (H\textsubscript{2}O\textsubscript{2}) with a 3:1 ratio (Piranha etch solution). The strongly oxidizing strength of this solution removes most organic matter from the pore walls while hydroxylating most surfaces (i.e., adding OH groups) in the process. The porous Vycor monoliths are immersed in the solution for 3 weeks and subsequently for 3 days in ultra pure water. After drying the process is repeated one more time. Before the experiment the samples are dried by introducing them in a vacuum chamber for 48 hours. Using pure water for every experiment avoids the necessity to repeat the whole cleaning process between different measurements, hence in this case only the drying is to be performed multiple times. . 

For the measurements we use a vertically oriented dilatometer (Linseis L75 VS500) with a precision up to \SI{0.1}{\micro\metre} and a temporal resolution of \SI{1}{\second}.

Humidity control during our experiments is achieved by using a custom-built humidity chamber that can be integrated into the  dilatometry setup. The desired relative humidity can be maintained by controlling the incoming air flow, which is a mixture of pure dry nitrogen and water vapor. The samples are exposed to the desired pre-imbibition humidity, where they remain until equilibrium is reached (monitored by dilatometry). After equilibration, the initially empty liquid reservoir is filled by injecting liquid through an external syringe until the water level reaches the sample, initiating the capillary filling process.

To seal the side facets of our samples, we used pressure-sensitive adhesive (PSA) tape provided by Tesa\textregistered. The adhesive tape consists of a polypropylene (PP) - polyethylene (PE) backing film for transparency and flexibility, paired with a polymeric adhesive layer. Both the adhesive and the liner are hydrophobic, which, combined with the viscoelastic properties of the tape (very low elastic modulus), makes it an excellent solution for sealing our specimens without compromising the mechanical stiffness of the monoliths. The adhesive mechanism of the PSA is based on the spreading of the polymer over the applied surface under pressure and the bonding of the polymers to the surface by van der Waals forces. Since the strain-driven pressures involved in the deformation process are on the order of \SI{}{\mega \pascal}, the effect of the sealing on the monolith deformation is neglected. The Tesa\textregistered \space tape used is optimized for easy removal with minimal residue. Using a magnifying glass, we have observed that small traces of polymer remain on the surface after the PSA is removed in between measurements. We therefore treat the surface with isopropanol to remove the remaining residues and to facilitate evaporation after washing the samples with ultrapure water.

\subsection{Molecular Dynamics Simulations}
Average quantities from the imbibition process are obtained from 20 individual Molecular Dynamics simulations with alternating starting velocities of the water molecules conducted with the Large-scale Atomic/Molecular Massively Parallel Simulator (LAMMPS) \cite{Thompson2022LAMMPSScales}. As host material, a periodically repeated amorphous silica pore initially introduced in Ref.~\cite{Ugliengo2008RealisticCalculations, DellePiane2014Large-scaleSystem} is used. The silica-silica interactions are modeled by a modified Demiralp potential \cite{Meiner2014ComputationalSilica}, water-silica interactions are modelled as described in Ref.~\cite{Cole2007DevelopmentBonding, Butenuth2012AbInterfaces}, and water-water interactions are modelled with the TIP4P/2005 water model \cite{Abascal2005ATIP4P/2005}. The simulated pore structure consists of amorphous silica $\mathrm{SiO_2}$. The surface of the porous structure is saturated with hydroxyl groups of \SI{7.23}{\per\nano\metre\squared} density. The total longitudinal pore length is around \SI{13}{\nano\metre} along the $z$ axis of the simulation box. However, in the simulation the water column starts imbibing from just inside the pore leaving around \SI{12}{\nano\metre} of possible imbibition length. In the lateral direction, that is the $x$ and $y$ direction of the simulation box, the pores are hexagonally shaped and have a diameter of around \SI{3}{\nano\metre} while the thickness of the pore walls is around \SI{1}{\nano\metre}. The imbibition process is simulated with periodic boundary conditions in all directions in an isothermal-isobaric ensemble (NpT) at \SI{1}{\bar} and \SI{300}{\kelvin} employing a Nosé–Hover barostat \cite{Martyna1994ConstantAlgorithms,Parrinello1981PolymorphicMethod,Coregliano2006AEnsemble,Shinoda2004RapidStress}. To trigger the imbibition process a water reservoir outside of the nanopores at the longitudinal pore ends is created. Due to the periodic boundary conditions the water can enter the pore from both sides. The $x$ and $y$ axes are coupled such that the rate of expansion or contraction of the simulation box is equal for both axes. The total simulated time is \SI{2000}{\pico\second} employing a \SI{1}{\femto\second} timestep (see Supporting information for further information).

The measured strains are defined as Cauchy strains $\epsilon(t) = [L(t) - L(t=t_0)] / L(t=t_0)$ where $L(t=t_0)$ constitutes the pore extensions at the start of the simulation.
The radial strain $\epsilon_{\perp}(t)$ is measured indirectly through the change of the simulation box extension $L_\mathrm{box}$ in the lateral direction.
In contrast the longitudinal strain $\epsilon_{\parallel}(t)$ is measured by the distances between the groups of atoms forming the last \SI{0.5}{\nano\metre} of bulk material of the pores on each end (see Supporting information for further information). The thermal motions of the atoms lead to strongly varying distances during the simulation. These fluctuations account for the high standard deviation in the measured longitudinal strain represented in the error bars in Fig.~\ref{fig:simulation_results}.

\acknow{Patrick Huber dedicates this work to
Klaus Knorr, Mainz, Germany, on the occasion of his 80th birthday, with immense gratitude for his mentorship in physics and for introducing him to the highly fascinating field of nanoporous
media. We are indebted to Klaus Knorr (Mainz, Germany) and Jörg Weißmüller (Hamburg, Germany) for a critical reading of the manuscript and for their helpful comments. J.S., P.H. and R.M. acknowledge funding by the Deutsche Forschungsmeinschaft (DFG, German Research Foundation) within the Research Training Group TRG 2462 ‘Processes in natural and technological Particle-Fluid-Systems (PintPFS)’ (Project No. 390794421) and within the Collaborative Research Centre CRC 986 'Tailor-Made Multi-Scale Materials Systems' (Project No.192346071). L.D. was supported by the Data Science in Hamburg HELMHOLTZ Graduate School DASHH, Helmholtz Association Grant-No. HIDSS-0002. L.G. profited from the DFG
research grant ”Dynamic Electrowetting at Nanoporous Surfaces: Switchable Spreading,
Imbibition, and Elastocapillarity”, Project number 422879465 (SPP 2171), and from the Cluster of Excellence ”Understanding Written Artefacts” (EXC 2176). This research was supported in part through the Maxwell computational resources operated at Deutsches Elektronen-Synchrotron DESY, Hamburg, Germany. M.F and P.H. thank the Free and Hanseatic City of Hamburg for funding within the research group 'Control of the special properties of water in nanopores'. We also acknowledge the scientific exchange and support of the Centre for Molecular Water Science CMWS, Hamburg (Germany).} 

\showacknow{}
%\bibliography{HuberLabReferences,LarsReferences}

\begin{thebibliography}{10}

\bibitem{Yao2013}
Yao X, et~al. (2013) {Adaptive fluid-infused porous films with tunable
  transparency and wettability}.
\newblock {\em Nature Materials} 12(6):529--534.

\bibitem{Wang2018FerrofluidInfused}
Wang W, et~al. (2018) {Multifunctional ferrofluid-infused surfaces with
  reconfigurable multiscale topography}.
\newblock {\em Nature} 559(7712):77--82.

\bibitem{Brinker2022}
Brinker M, Huber P (2022) {Wafer‐Scale Electroactive Nanoporous Silicon:
  Large and Fully Reversible Electrochemo‐Mechanical Actuation in Aqueous
  Electrolytes}.
\newblock {\em Advanced Materials} 34:2105923.

\bibitem{Sentker2019}
Sentker K, et~al. (2019) {Self-assembly of liquid crystals in nanoporous solids
  for adaptive photonic metamaterials}.
\newblock {\em Nanoscale} 11(48):23304--23317.

\bibitem{Cencha2020}
Cencha LG, Dittrich G, Huber P, Berli CLA, Urteaga R (2020) {Precursor Film
  Spreading during Liquid Imbibition in Nanoporous Photonic Crystals}.
\newblock {\em Physical Review Letters} 125(23):234502.

\bibitem{Emmerich2022}
Emmerich T, et~al. (2022) {Enhanced nanofluidic transport in activated carbon
  nanoconduits}.
\newblock {\em Nature Materials} pp. 1--8.

\bibitem{Simon2008}
Simon P, Gogotsi Y (2008) {Materials for electrochemical capacitors}.
\newblock {\em Nature Materials} 7:845--854.

\bibitem{Zhang2021a}
Zhang J, Chen B, Chen X, Hou X (2021) {Liquid‐Based Adaptive Structural
  Materials}.
\newblock {\em Advanced Materials} 2005664.

\bibitem{Wong2011}
Wong TS, et~al. (2011) {Bioinspired self-repairing slippery surfaces with
  pressure-stable omniphobicity}.
\newblock {\em Nature} 477(7365):443--447.

\bibitem{Meehan1927expansioncharcoal}
Meehan FT, Hardy WB (1927) {The expansion of charcoal on sorption of carbon
  dioxide}.
\newblock {\em Proceedings of the Royal Society of London. Series A, Containing
  Papers of a Mathematical and Physical Character} 115(770):199--207.

\bibitem{Bangham1928}
Bangham DH, Fakhoury N (1928) {The expansion of charcoal acompanying sorption
  of gases and vapours}.
\newblock {\em Nature} 122:681--682.

\bibitem{Scherer1986DilatationGlass}
Scherer GW (1986) {Dilatation of Porous Glass}.
\newblock {\em Journal of the American Ceramic Society} 69(6):473--480.

\bibitem{Prass2009}
Prass J, Mueter D, Fratzl P, Paris O (2009) {Capillarity-driven deformation of
  ordered nanoporous silica}.
\newblock {\em Applied Physics Letters} 95(8):83121.

\bibitem{Gor2010}
Gor GY, Neimark AV (2010) {Adsorption-Induced Deformation of Mesoporous
  Solids}.
\newblock {\em Langmuir} 26(16):13021--13027.

\bibitem{Schappert2014VycorDeformation}
Schappert K, Pelster R (2014) {Unexpected Sorption-Induced Deformation of
  Nanoporous Glass: Evidence for Spatial Rearrangement of Adsorbed Argon}.
\newblock {\em Langmuir} 30(46):14004--14013.

\bibitem{Balzer2014}
Balzer C, et~al. (2014) {Relationship between pore structure and
  sorption-induced deformation in hierarchical silica-based monoliths}.
\newblock {\em Zeitschrift fuer Physikalische Chemie} 229:1189--1209.

\bibitem{Grosman2015}
Grosman A, Puibasset J, Rolley E (2015) {Adsorption-induced strain of a
  nanoscale silicon honeycomb}.
\newblock {\em EPL (Europhysics Letters)} 109(5):56002.

\bibitem{Gor2017}
Gor GY, Huber P, Bernstein N (2017) {Adsorption-induced deformation of
  nanoporous materials—A review}.
\newblock {\em Applied Physics Reviews} 4(1):011303.

\bibitem{Chen2019SorptionDeformation}
Chen M, Coasne B, Guyer R, Derome D, Carmeliet J (2019) {Molecular Simulation
  of Sorption-Induced Deformation in Atomistic Nanoporous Materials}.
\newblock {\em Langmuir} 35(24):7751--7758.

\bibitem{Yang2020}
Yang Q, et~al. (2020) {Capillary condensation under atomic-scale confinement}.
\newblock {\em Nature} 588(7837):250--253.

\bibitem{Harrellson2023}
Harrellson SG, et~al. (2023) {Hydration solids}.
\newblock {\em Nature} 619(7970):500--505.

\bibitem{Gor2024}
Gor GY, et~al. (2024) { Bacterial spores respond to humidity similarly to hydrogelss}.
\newblock {\em PNAS}  121 (10)e2320763121.

\bibitem{Gor2015}
Gor GY, et~al. (2015) {Elastic response of mesoporous silicon to capillary
  pressures in the pores}.
\newblock {\em Applied Physics Letters} 106:1--13.

\bibitem{Gor2016RevisitingStress}
Gor GY, Bernstein N (2016) {Revisiting Bangham's law of adsorption-induced
  deformation: changes of surface energy and surface stress}.
\newblock {\em Physical Chemistry Chemical Physics} 18(14):9788--9798.

\bibitem{Gor2015ElasticPores}
Gor GY, et~al. (2015) {Elastic response of mesoporous silicon to capillary
  pressures in the pores}.
\newblock {\em Applied Physics Letters} 106(26):261901.

\bibitem{duprat_aristoff_stone_2011}
Duprat C, Aristoff J, Stone HA (2011) {Dynamics of elastocapillary rise}.
\newblock {\em Journal of Fluid Mechanics} 679:641–654.

\bibitem{Weijs2013Elasto-capillaritySolids}
Weijs JH, Andreotti B, Snoeijer JH (2013) {Elasto-capillarity at the nanoscale:
  On the coupling between elasticity and surface energy in soft solids}.
\newblock {\em Soft Matter} 9(35):8494--8503.

\bibitem{Tortora2021}
Tortora M, et~al. (2021) {Giant Negative Compressibility by Liquid Intrusion
  into Superhydrophobic Flexible Nanoporous Frameworks}.
\newblock {\em Nano Letters} 21(7):2848--2853.

\bibitem{Michel2022DeformationHydrophobicMedium}
Michel L, et~al. (2022) {Bowtie-Shaped Deformation Isotherm of Superhydrophobic
  Cylindrical Mesopores}.
\newblock {\em Langmuir} 38(1):211--220.

\bibitem{Hoberg2014}
Hoberg TB, Verneuil E, Hosoi AE (2014) {Elastocapillary flows in flexible
  tubes}.
\newblock {\em Physics of Fluids} 26(12):122103.

\bibitem{Ha2018Poro-elasto-capillarySponges}
Ha J, et~al. (2018) {Poro-elasto-capillary wicking of cellulose sponges}.
\newblock {\em Science Advances} 4(3):1--7.

\bibitem{Siddique2009CapRiseDefPorousMedium}
Siddique JI, Anderson DM, Bondarev A (2009) {Capillary rise of a liquid into a
  deformable porous material}.
\newblock {\em Physics of Fluids} 21(1):013106.

\bibitem{Levitz1991}
Levitz P, Ehret G, Sinha SK, Drake JM (1991) {Porous Vycor glass: The
  microstructure as probed by electron microscopy, direct energy transfer,
  small-angle scattering, and molecular adsorption}.
\newblock {\em J. Chem. Phys.} 95:6151.

\bibitem{Huber1999}
Huber P, Knorr K (1999) {Adsorption-desorption isotherms and x-ray diffraction
  of Ar condensed into a porous glass matrix}.
\newblock {\em Phys. Rev. B} 60(18):12657.

\bibitem{Levitz1998}
Levitz P (1998) {Off-lattice reconstruction of porous media: Critical
  evaluation, geometrical confinement and molecular transport}.
\newblock {\em Advances in Colloid and Interface Science}.

\bibitem{Gommes2018StochasticModelsPorousMedia}
Gommes CJ (2018) {Stochastic models of disordered mesoporous materials for
  small-angle scattering analysis and more}.
\newblock {\em Microporous and Mesoporous Materials} 257:62--78.

\bibitem{Gruener2009CapillaryNanopores}
Gruener S, Hofmann T, Wallacher D, Kityk AV, Huber P (2009) {Capillary rise of
  water in hydrophilic nanopores}.
\newblock {\em Physical Review E - Statistical, Nonlinear, and Soft Matter
  Physics} 79(6):067301.

\bibitem{Gruener2012}
Gruener S, et~al. (2012) {Anomalous front broadening during spontaneous
  imbibition in a matrix with elongated pores}.
\newblock {\em Proceedings of the National Academy of Sciences of the United
  States of America} 109(26):10245--50.

\bibitem{Gruener2015}
Gruener S, Hermes HE, Schillinger B, Egelhaaf SU, Huber P (2015) {Capillary
  rise dynamics of liquid hydrocarbons in mesoporous silica as explored by
  gravimetry, optical and neutron imaging: Nano-rheology and determination of
  pore size distributions from the shape of imbibition fronts}.
\newblock {\em Colloids and Surfaces A: Physicochemical and Engineering
  Aspects} 496:13--27.

\bibitem{Page1993}
Page JH, Liu J, Abeles B, Deckman HW, Weitz DA (1993) {Pore-Space Correlations
  in Capillary Condensation in Vycor}.
\newblock {\em Phys. Rev. Lett.} 71:1216--1219.

\bibitem{Amberg1952b}
Amberg CH, McIntosh R (1952) {a Study of Adsorption Hysteresis By Means of
  Length Changes of a Rod of Porous Glass}.
\newblock {\em Canadian Journal of Chemistry} 30(12):1012--1032.

\bibitem{Cencha2019}
 Cencha LG,  Huber P, Kappl M, Floudas G, Steinhart M, Berli CLA, Urteaga R (2019) {Nondestructive high-throughput screening of nanopore geometry in porous membranes by imbibition}.
\newblock {\em Appl. Phys. Lett.} 115 (11): 113701.

\bibitem{Gruener2011}
Gruener S, Huber P (2011) {Imbibition in mesoporous silica: rheological
  concepts and experiments on water and a liquid crystal.}
\newblock {\em J. Phys.: Condens. Matter} 23(18):184109.

\bibitem{Schlaich2017HydrationFriction}
Schlaich A, Kappler J, Netz RR (2017) {Hydration Friction in Nanoconfinement:
  From Bulk via Interfacial to Dry Friction}.
\newblock {\em Nano Letters} 17(10):5969--5975.

\bibitem{Maillet2022}
Maillet B, Dittrich G, Huber P, Coussot P (2022) {Diffusionlike Drying of a
  Nanoporous Solid as Revealed by Magnetic Resonance Imaging}.
\newblock {\em Physical Review Applied} 10(1):1.

\bibitem{Seker2008WettingEvaporation}
Seker E, Begley MR, Reed ML, Utz M (2008) {Kinetics of capillary wetting in
  nanoporous films in the presence of surface evaporation}.
\newblock {\em Applied Physics Letters} 92(1):013128.

\bibitem{Wheeler2008}
Wheeler TD, Stroock AD (2008) {The transpiration of water at negative pressures
  in a synthetic tree}.
\newblock {\em Nature} 455:208.

\bibitem{Camplisson2015}
Camplisson CK, Schilling KM, Pedrotti WL, Stone HA, Martinez AW (2015) {Two-ply
  channels for faster wicking in paper-based microfluidic devices}.
\newblock {\em Lab on a Chip} 15(23):4461--4466.

\bibitem{Huang2021}
Huang Z, et~al. (2021) {Fast Water Evaporation from Nanopores}.
\newblock {\em Advanced Materials Interfaces} 8(14).

\bibitem{Amberg1952}
Amberg CH, Mcinthosh R (1952) {A Study of Adsorption Hysteresis By Means of
  Length Changes of A Rod of Porous Glass}.
\newblock {\em Canadian Journal of Chemistry-revue Canadienne De Chimie}
  30(12):1012--1032.

\bibitem{Weissmuller2010DeformationForces}
Weissm{\"{u}}ller J, Duan HL, Farkas D (2010) {Deformation of solids with
  nanoscale pores by the action of capillary forces}.
\newblock {\em Acta Materialia} 58(1):1--13.

\bibitem{Gor2010a}
Gor GY, Neimark AV (2010) {Adsorption-induced deformation of mesoporous
  solids.}
\newblock {\em Langmuir : the ACS journal of surfaces and colloids}
  26(16):13021--13027.

\bibitem{Gennes2004}
de~Gennes PG, Brochard-Wyart F, Quere D (2004) {\em {Capillarity and Wetting
  Phenomena: Drops, Bubbles, Pearls, Waves}}.
\newblock (Springer, New York).

\bibitem{Brunauer1956}
Brunauer S, Kantro DL, Weise CH (1956) {The surface energies of amorphous
  silica and hydrous amorphous silica}.
\newblock {\em Canadian Journal of Chemistry} 34(10):1483--1496.

\bibitem{Diaz2022}
D{\'{i}}az D, et~al. (2022) {How water wets and self-hydrophilizes nanopatterns
  of physisorbed hydrocarbons}.
\newblock {\em Journal of Colloid and Interface Science} 606:57--66.

\bibitem{Ugliengo2008RealisticCalculations}
Ugliengo P, et~al. (2008) {Realistic Models of Hydroxylated Amorphous Silica
  Surfaces and MCM-41 Mesoporous Material Simulated by Large-scale Periodic
  B3LYP Calculations}.
\newblock {\em Advanced Materials} 20(23):4579--4583.

\bibitem{DellePiane2014Large-scaleSystem}
Delle~Piane M, Corno M, Pedone A, Dovesi R, Ugliengo P (2014) {Large-scale
  B3LYP simulations of ibuprofen adsorbed in MCM-41 mesoporous silica as drug
  delivery system}.
\newblock {\em Journal of Physical Chemistry C} 118(46):26737--26749.

\bibitem{Dimitrov2007}
Dimitrov DI, Milchev A, Binder K (2007) {Capillary rise in nanopores: Molecular
  dynamics evidence for the Lucas-Washburn equation}.
\newblock {\em Phys. Rev. Lett.} 99:54501.

\bibitem{Thompson2022LAMMPSScales}
Thompson AP, et~al. (2022) {LAMMPS - a flexible simulation tool for
  particle-based materials modeling at the atomic, meso, and continuum scales}.
\newblock {\em Computer Physics Communications} 271:108171.

\bibitem{Meiner2014ComputationalSilica}
Mei{\ss}ner RH, Schneider J, Schiffels P, Colombi~Ciacchi L (2014)
  {Computational prediction of circular dichroism spectra and quantification of
  helicity loss upon peptide adsorption on silica}.
\newblock {\em Langmuir} 30(12):3487--3494.

\bibitem{Cole2007DevelopmentBonding}
Cole DJ, Payne MC, Cs{\'{a}}nyi G, Spearing SM, Ciacchi LC (2007) {Development
  of a classical force field for the oxidized Si surface: Application to
  hydrophilic wafer bonding}.
\newblock {\em Journal of Chemical Physics} 127(20):204704.

\bibitem{Butenuth2012AbInterfaces}
Butenuth A, et~al. (2012) {Ab initio derived force-field parameters for
  molecular dynamics simulations of deprotonated amorphous-SiO2/water
  interfaces}.
\newblock {\em physica status solidi (b)} 249(2):292--305.

\bibitem{Abascal2005ATIP4P/2005}
Abascal JL, Vega C (2005) {A general purpose model for the condensed phases of
  water: TIP4P/2005}.
\newblock {\em The Journal of Chemical Physics} 123(23):234505.

\bibitem{Martyna1994ConstantAlgorithms}
Martyna GJ, Tobias DJ, Klein ML (1994) {Constant pressure molecular dynamics
  algorithms}.
\newblock {\em The Journal of Chemical Physics} 101(5):4177--4189.

\bibitem{Parrinello1981PolymorphicMethod}
Parrinello M, Rahman A (1981) {Polymorphic transitions in single crystals: A
  new molecular dynamics method}.
\newblock {\em Journal of Applied Physics} 52(12):7182--7190.

\bibitem{Coregliano2006AEnsemble}
Coregliano LN, et~al. (2006) {A Liouville-operator derived measure-preserving
  integrator for molecular dynamics simulations in the isothermal–isobaric
  ensemble}.
\newblock {\em Journal of Physics A: Mathematical and General} 39(19):5629.

\bibitem{Shinoda2004RapidStress}
Shinoda W, Shiga M, Mikami M (2004) {Rapid estimation of elastic constants by
  molecular dynamics simulation under constant stress}.
\newblock {\em Physical Review B} 69(13):134103.

\bibitem{SanchezCalzado2024}
Sanchez Calzado J (2004) {Deformation dynamics of nanopores upon water imbibition: Optical,
gravimetric and dilatometry experiments}.
TORE. https://doi.org/10.15480/882.13233. Deposited 21 August 2024.

\bibitem{SanchezCalzado2024}
Sanchez Calzado J (2024) {Deformation dynamics of nanopores upon water imbibition: Optical, gravimetric and dilatometry experiments}.
TORE. https://doi.org/10.15480/882.13233. Deposited 21 August 2024.

\bibitem{DammannL2024}
Dammann L (2024) {Deformation dynamics of nanopores upon water imbibition:  Molecular dynamics simulationss}. TORE. https://doi.org/10.15480/882.13223. Deposited 21 August 2024.

\end{thebibliography}

\end{document}